\documentclass[conference]{IEEEtran}
\pdfoutput=1
\usepackage{cite}
\usepackage{amsmath,amssymb,amsfonts}
\usepackage{algorithmic}
\usepackage{graphicx}
\usepackage{textcomp}
\usepackage[table]{xcolor}
\usepackage{tikz}
\usepackage{pgfplots}
\usetikzlibrary{3d, calc}
\usetikzlibrary{shapes.geometric, arrows,positioning,decorations.pathreplacing}
\usepackage{url}
\usepackage{comment}
\pgfplotsset{compat=1.17}
\usepackage{tabularx}

\newtheorem{Remark}{Remark}

\usepackage{booktabs}
\usepackage{multirow}
\usepackage{numprint}
\npstyleenglish

\def\BibTeX{{\rm B\kern-.05em{\sc i\kern-.025em b}\kern-.08em
T\kern-.1667em\lower.7ex\hbox{E}\kern-.125emX}}

\begin{document}
	
	\title{Collision Risk Analysis for LEO Satellites with Confidential Orbital Data}
	
	\author{\IEEEauthorblockN{Svenja Lage, Felicitas Hörmann}
	\IEEEauthorblockA{\textit{Communications and Navigation} \\
	\textit{German Aerospace Center (DLR)}\\
	Oberpfaffenhofen, Germany\\
	\hspace{-0.7cm}\texttt{\{svenja.lage, felicitas.hoermann\}@dlr.de}}
	\and
	\IEEEauthorblockN{Felix Hanke, Michael Karl}
	\IEEEauthorblockA{\textit{AI Safety and Security} \\
	\textit{German Aerospace Center (DLR)}\\
	Sankt Augustin, Germany\\
	\texttt{\{felix.hanke, michael.karl\}@dlr.de}}
	}
 
	\maketitle
 
	\begin{abstract}
		The growing number of satellites in low Earth orbit (LEO) has increased concerns about the risk of satellite collisions, which can ultimately result in the irretrievable loss of satellites and a growing amount of space debris. To mitigate this risk, accurate collision risk analysis is essential. However, this requires access to sensitive orbital data, which satellite operators are often unwilling to share due to privacy concerns.
		This contribution proposes a solution based on fully homomorphic encryption (FHE) and thus enables secure and private collision risk analysis. In contrast to existing methods, this approach ensures that collision risk analysis can be performed on sensitive orbital data without revealing it to other parties.
		To display the challenges and opportunities of FHE in this context, an implementation of the CKKS scheme is adapted and analyzed for its capacity to satisfy the theoretical requirements of precision and run time.
	\end{abstract}

	\section{Introduction}
	
	Large-scale satellite projects like the Starlink network, which has launched over \numprint{6000} low Earth orbit (LEO) satellites within the past six years \cite{StarlinkStat}, have significant implications for the space environment and highly increase the probability of satellite collisions. For the increasingly critical analysis of collision risk, access to precise orbital data is essential to obtain robust predictions. However, due to privacy concerns, operators are often reluctant to share detailed data, which necessitates the use of less accurate observational data instead. To address this issue, we explore the usage of fully homomorphic encryption (FHE) for precise satellite collision prediction.
	FHE is an advanced cryptographic technique that enables the computation on encrypted data without the need for decryption and re-encryption and thus provides continuous privacy for the underlying information. More precisely, a computation on FHE-encrypted ciphertexts yields a result, whose decryption equals the result of the same computation on the respective plaintexts. In our proposed scenario, operators share their encrypted orbital data, allowing for the collision calculation to be performed without disclosing sensitive information to any other party. Consequently, operators can benefit from more accurate collision predictions while preserving the confidentiality of their data. \\[-.6em]
 
	We first provide a comprehensive overview of the mathematical foundations underlying collision risk analysis, examining the existing process and its attendant requirements. In order to develop a satellite collision risk model based on FHE, we present an introduction to FHE in the single-party scenario, highlighting the pivotal role of the bootstrapping procedure. This procedure, while enabling homomorphic schemes for the first time, also constitutes a significant limiting factor with respect to computational complexity. \\[-.6em]
	
	We then introduce the two basic concepts to expand a fully homomorphic scheme to the multi-party case, namely we study multi-key and threshold FHE. Although threshold FHE is more prevalent, both approaches possess diverse advantages in the context of our specific application. Furthermore, we provide a concrete exemplar of the challenges and opportunities inherent in our approach in Section \ref{Fully_Homomorphic_Encryption_for_Satellites} utilizing the CKKS scheme \cite{CKKS}. Despite being associated with several hurdles, the analysis underscores the potential of an FHE approach in the context of satellite collision risk calculation, thereby emphasizing the necessity of continued advancements in this research domain.

	\section{Background on Collision Risk Assessment}\label{BackgroundCollisionRisk}
	
	This section introduces the principles of collision probability calculation based on \cite{chan2008spacecraft}, \cite{Aida} and \cite{CloseApproach}. Consider two satellites, denoted by $s_1$ and $s_2$. Each satellite is modeled as a three-dimensional spherical object, with radii $r_1$ and $r_2$ for $s_1$ and $s_2$, respectively. The two satellites collide whenever the two spheres overlap. For a fixed time $t>0$, let $\mu_1, \mu_2\in\mathbb{R}^3$ denote the estimated position of the satellites $s_1$ and $s_2$ respectively in the Cartesian coordinate system. Due to external forces such as atmospheric influence, even the satellite operators can only estimate the position at a given time $t$ in the future with with a certain error. Since the exact structure of the error is unknown, we model the distribution of the real physical location of each satellite by a normal distribution $p_i\sim N(\mu_i,C_i)$ for a given covariance matrix $C_i\in\mathbb{R}^{3\times 3}$ such that the probability density function is given by
	\begin{align*}
		g_{\mu_i,C_i}(x)= &\tfrac{1}{\sqrt{(2\pi)^3\det(C_i)}}\cdot
		\exp\left(-\tfrac{1}{2}(x-\mu_i)^TC_i^{-1}(x-\mu_i)\right)
	\end{align*}
	for $x\in\mathbb{R}^3$ and $i\in\{1,2\}$.
	To simplify the calculation, it is common to attribute all mass towards one of the objects, whereas the second object is considered a point particle with combined positional uncertainty. Although both space objects are interchangeable, we go along with the usual convention and consider satellite $s_1$ as an object with radius $r=r_1+r_2$, but no positional uncertainty. Consequently, satellite $s_2$, which we place in the origin, is considered as a point particle and its position is distributed as $\widetilde{p_2}\sim N(0,C_1+C_2)$. Note that for simplicity, we assume that the positional errors are uncorrelated. However, we acknowledge that certain factors, such as drags, indeed induce correlations between positional errors but as discussed in \cite[Section 2.6]{book}, these have a negligible impact on the overall analysis.\\[-.6em]
	
	As the first object is moving through the combined covariance ellipsoid, a collision occurs at time $t$ with a certain probability calculated by
	\begin{align}
		\label{3DIntegral}
		P_{col}(t)= \int_{S_r} g_{0,C_1+C_2}(x) \; dx,
	\end{align}
	where $S_r$ is the sphere of radius $r$ spanned by the satellite $s_1$ around its relative position at time $t$. Note that not only $S_r$ but also $C_1+C_2$ depend on $t$ but we omit this dependency for readability. The probability that the two satellites collide within a given time period $[t_1,t_2]$ is finally given by
	\begin{align}
		\label{P_collision}
		\int_{t_1}^{t_2} P_{col}(t)\; dt.
	\end{align}
	Solving this integral using Monte-Carlo simulations requires a large number of samples, resulting in a computationally expensive process (compare \cite{MonteCarlo1} or \cite{MonteCarlo2}). Although there exist other analytical and numerical approaches to deal with the integral (\ref{P_collision}) (see \cite{Review} for an overview), none of these simultaneously meet the requirements of precision and computational speed in a broad range of scenarios. \\[-.6em]
	
	Hence, in recent research, collision risk analysts often differentiate between two types of scenarios: the short-term encounter scenario, which refers to a situation where the two objects have a high relative velocity and a brief approach (often lasting only a few seconds), and the long-encounter scenario, in which the relative velocity is lower and the encounter duration exceeds a few seconds. Both scenarios exhibit noticeable differences in their behavior \cite{book} and are thus the focus of separate research efforts. Owing to the particularly increasing number of orbital objects in the LEO, our attention is focused on the short-term encounter scenario.
	In this particular scenario, the encounter time is small such that, within this period, the normally curved motion of the objects can be approximated with a linear motion with a relatively small error margin. Furthermore, we can assume that the collision probability $P_{col}$ is constant over the short encounter period which simplifies (\ref{P_collision}) to the three-dimensional case in (\ref{3DIntegral}). To preclude underestimation of the collision probability, the parameters are fixed at the time of closest approach (TCA).\\[-.6em]
 
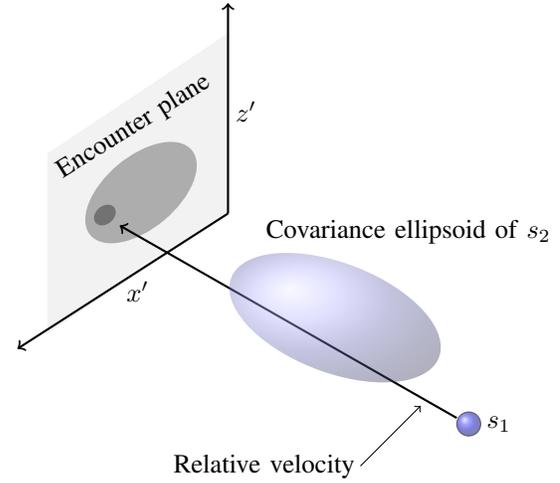
\begin{figure}
		\centering
		\begin{tikzpicture}[scale=0.8]
			\definecolor{darkgray}{gray}{0.3}
			\coordinate (O) at (0,0,0); %
			\coordinate (j) at (3,2,0); %
			\coordinate (k) at (0,3,0); %
			\coordinate (i) at (0,2,3); %
			\coordinate (B) at (7,-1.5,0);
			\coordinate (plane) at (4.3,4.3);
			\fill[gray!20, opacity=0.5] (O) -- (j) -- ($(j)+(k)$) -- (k) -- cycle;
			\begin{scope}[rotate=40]
				\fill[color=gray, opacity=0.6] (2.7,0.8) ellipse (1.1 and 0.6);
			\end{scope}
			\begin{scope}[rotate=40]
				\fill[color=black!90, opacity=0.4] (2,0.9) ellipse (0.2 and 0.15);
			\end{scope}
			\draw[->,thick] (j) -- (-0.5,-0.25,0);
			\draw[->,thick] (j) -- (3,5.5,0);
			\draw[->,thick] (B)+(-0.2,0.1,0) -- (1.2,1.8,0);
			\draw (7.5,-1.5,0) node {$s_1$};
			\draw (6,1.7) node {Covariance ellipsoid of $s_2$};
			\draw (3.6,-2.2) node {Relative velocity};
			\draw (1.5,0.7,0) node {$x'$};
			\draw (3.3,3.7,0) node {$z'$};
			\draw[->] (5.2,-2.2) -- (6.2,-1.2);
			\node [rotate=32,left=1.2cm of plane] {Encounter plane};
			\draw[ball color = blue, opacity=0.5, shading=ball] (B) circle (0.2cm);
			\begin{scope}
				[canvas is zx plane at y=1, rotate around={40:(0,0)}]
				\shade[ball color=blue!40, opacity=0.4] (5,3) ellipse (3.5 and 1.25);
			\end{scope}
		\end{tikzpicture}
		\caption{Representation of satellite $s_1$ as sphere with combined radius $r=r_1+r_2$ and satellite $s_2$ as point particle with combined covariance matrix. Additionally, the encounter plane normal to the relative velocity as well as the projection of $s_1$ and $s_2$ onto the plane are displayed. Inspired by \cite{CloseApproach}.}
		\label{Fig1}
	\end{figure}

	To further simplify the calculation, we define a coordinate system with respect to the encounter plane. Therefore, let the $y'$-axis be along the relative velocity vector $v=v_1-v_2$ and choose the $(x',z')$-plane -- the so-called encounter plane -- normal to $v$. By doing so, the distance between the two satellites is purely based on their distance in the $(x',z')$-plane such that the collision probability can be described via the projections of the objects onto the encounter plane (compare Figure \ref{Fig1}). As a result,
	\begin{align*}
		P_{col}&= \int_{B_r} \frac{1}{2\pi \sigma_{x'}\sigma_{z'}} e^{-\frac{1}{2}\left[\left(\frac{x'}{\sigma_{x'}}\right)^2+\left(\frac{z'}{\sigma_{z'}}\right)^2\right]} \; dx' \; dz',
	\end{align*}
	where for simplicity we assume that the $x'$- and $z'$-axes are chosen such that the covariance matrix is diagonal with elements $\sigma_{x'}^2$ and $\sigma_{z'}^2$ and $B_r$ denotes the cross section of the two-dimensional projection in the encounter plane. \\[-.6em]
	
	In 1992, Foster and Estes \cite{Foster} assumed a circular cross-section, allowing them to express the integral in polar coordinates as
	\begin{align}
		\label{Approx}
		P_{col}&= \frac{1}{2\pi \sigma_{x'}\sigma_{z'}} \int_0^{r}\int_{0}^{2\pi} ye^{-\frac{1}{2}y^2 \left(\frac{\cos(\phi)^2}{\sigma_{x'}^2}+\frac{\sin(\phi)^2}{\sigma_{z'}^2}\right)} d\phi \, dy \notag \\
		&=: \int_0^{r}\int_{0}^{2\pi} p(y,\phi) \; d\phi \; dy,
	\end{align}
	where for simplicity we assume that $B_r$ is centered at the origin. Since this model is still in use by the NASA \cite{nasa} as well as by the German Aerospace Center \cite{Aida}, our model will be primarily based on this calculation although other approximations exist and may stimulate further research.
	In certain scenarios, it may be advantageous to assume a constant probability function $p$ within the region of integration, thereby simplifying the integral (\ref{Approx}) to a straightforward evaluation of $p$ without the necessity of numerical integration. However, as Aida et al. \cite{Aida} noted, this approximation introduces a non-negligible error when $r$ is large and the combined covariance is small.\\[-.6em]
	
	It is well-known that no closed-form expression exists for the solution of the integral (\ref{Approx}), necessitating its numerical solution. We approximate both integrals with a numerical integration rule using a fixed stepsize $h_r$ in the radius and a fixed stepsize $h_{\phi}$ in the angle. The sampling points are given by $(y_i, \phi_j)$ where $y_i = i \cdot h_r$ and $\phi_j = j \cdot h_{\phi}$ for $i = 1, \ldots, N=\lfloor \frac{r}{h_r}\rfloor$ and $j = 1, \ldots, M=\lfloor\frac{2\pi}{h_{\phi}}\rfloor$. Although algorithms with adaptive stepsizes are generally more efficient and reduce computational cost, their use is limited in this particular context due to the encryption of the data. Therefore, a fixed stepsize is employed for the numerical integration. We test different approaches for the numerical integration:\\[-.6em]
	
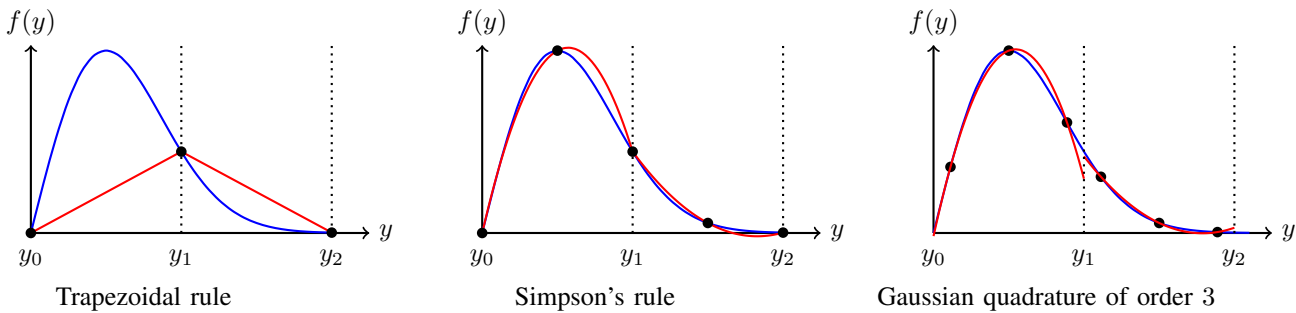
\begin{figure*}
		\begin{tikzpicture}[scale=1]
			\begin{scope}
				[shift={(0,0)}]
				\draw[thick,->] (0,0) -- (4.5,0) node[right] {\(y\)};
				\draw[thick,->] (0,0) -- (0,2.5) node[above] {\(f(y)\)};
				\draw[domain=0:4,smooth,variable=\x,thick,blue] plot ({\x},{4*\x*pow(e,-0.5*(pow(\x,2)))});
				
				\draw[thick,red] (0 ,0) -- (2,4*0.27067);
				\draw[thick,red] (2,4*0.27067) -- (4,4*0.00134);
				
				\foreach \x in {0,2,4} {
					\fill[black] ({\x},{4*\x*pow(e,-0.5*(pow(\x,2)))}) circle(2pt);
				}
				\draw[dotted,thick] (2,0) -- (2,2.5);
				\draw[dotted,thick] (4,0) -- (4,2.5);
    
				\node[below] at (2,-0.1) {\(y_1\)};
				\node[below] at (4,-0.1) {\(y_2\)};
				\node[below] at (0,-0.1) {\(y_0\)};
				\node[below] at (1.5,-0.6) {Trapezoidal rule};
			\end{scope}
   
			\begin{scope}
				[shift={(6,0)}]
				\draw[thick,->] (0,0) -- (4.5,0) node[right] {\(y\)};
				\draw[thick,->] (0,0) -- (0,2.5) node[above] {\(f(y)\)};
				\draw[domain=0:4,smooth,variable=\x,thick,blue] plot ({\x},{4*\x*pow(e,-0.5*(pow(\x,2)))});
    
				\draw[domain=0:2,smooth,variable=\x,thick,red] plot ({\x},{-4*0.471196*\x*\x+4*1.07773*\x});
				\draw[domain=2:4,smooth,variable=\x,thick,red] plot ({\x},{4*0.102679*\x*\x-4*0.750741*\x+4*1.36144});
    
				\foreach \x in {0,1,2,3,4} {
					\fill[black] ({\x},{4*\x*pow(e,-0.5*(pow(\x,2)))}) circle(2pt);}
     
				\draw[dotted,thick] (2,0) -- (2,2.5);
				\draw[dotted,thick] (4,0) -- (4,2.5);
    
				\node[below] at (2,-0.1) {\(y_1\)};
				\node[below] at (4,-0.1) {\(y_2\)};
				\node[below] at (0,-0.1) {\(y_0\)};
				\node[below] at (1.5,-0.6) {Simpson's rule};
			\end{scope}
			
			\begin{scope}
				[shift={(12,0)}]
				\draw[thick,->] (0,0) -- (4.5,0) node[right] {\(y\)};
				\draw[thick,->] (0,0) -- (0,2.5) node[above] {\(f(y)\)};
				\draw[domain=0:4.2,smooth,variable=\x,thick,blue] plot ({\x},{4*\x*pow(e,-0.5*(pow(\x,2)))});
				
				\foreach \x in {0.2254,1,1.7745,2.2254,3,3.7745} {
					\fill[black] ({\x},{4*\x*pow(e,-0.5*(pow(\x,2)))}) circle(2pt);
				}
				
				\draw[domain=0:2,smooth,variable=\x,thick,red] plot ({\x},{-4*0.521531*\x*\x+4*1.13842*\x-4*0.0103585});
				\draw[domain=2:4,smooth,variable=\x,thick,red] plot ({\x},{4*0.102885*\x*\x-4*0.736098*\x+4*1.31565});
				\draw[dotted,thick] (2,0) -- (2,2.5);
				\draw[dotted,thick] (4,0) -- (4,2.5);
				
				\node[below] at (2,-0.1) {\(y_1\)};
				\node[below] at (4,-0.1) {\(y_2\)};
				\node[below] at (0,-0.1) {\(y_0\)};
				\node[below] at (1.5,-0.6) {Gaussian quadrature of order $3$};
			\end{scope}
		\end{tikzpicture}
		\caption{Illustration of the Trapezoidal rule, the Simpson's rule and the Gaussian quadrature of order 3 (red) for the function $y\mapsto f(y)=ye^{-\frac{1}{2}y^2}$ (blue). The marked points are used to calculate the approximation.}
		\label{NumIntDarstellung}
	\end{figure*}

	\textbf{Trapezoidal rule:}
	The Trapezoidal rule employs a linear interpolation between two evaluation points, thereby generating a trapezoidal approximation, as illustrated in Figure \ref{NumIntDarstellung}. By selecting the interval bounds as the evaluation points, this rule minimizes the number of function evaluations required, rendering it a computationally efficient approach. However, this advantage comes at the expense of accuracy, as the resulting approximation is generally less precise compared to other numerical integration methods. \\[-.6em]
	
	\textbf{Trapezoidal and Simpson rule:}
	In \cite{Chan}, the authors suggested a hybrid approach to approximate he double integral, wherein the outer integral is approximated using the Trapezoidal rule, while the inner integral is approximated via Simpson's rule. In contrast to the Trapezoidal rule, Simpson's rule enables a more accurate approximation by utilizing a quadratic polynomial to interpolate between two sampling points, as illustrated in Figure \ref{NumIntDarstellung}. The quadratic fit is computed using the boundary points of the interval, as well as an additional evaluation point situated at the midpoint of the interval, thereby providing a more nuanced representation of the underlying function. \\[-.6em]
	
	\textbf{Gaussian quadrature rule:}
	In contrast to the previous methods, the Gaussian quadrature of order $n$ entails the selection of $n$ evaluation points situated within the interval boundaries. The integral is approximated by a weighted sum of the function values at these points. The determination of the evaluation points and their corresponding weights is facilitated through the utilization of Legendre polynomials, which ensures that the quadrature rule integrates polynomials of order $2n-1$ or lower exactly.\\[-.6em]
 
	\section{Current Status and Challenges}
 
	The collision risk analysis relies on accurate knowledge of the satellites' movement, necessitating the forecast of parameters $\mu_i$, $C_i$, and $r_i$ for $i\in\{1,2\}$ across the time period of interest. While the satellite radius is typically not a sensitive parameter, due to privacy concerns -- especially in the military setting -- operators often hesitate to share precise trajectory data. In practice, orbit prediction is based on tracking data provided by organizations such as the 19th Space Defense Squadron (19th SDS), operated by the United States Space Force. The SDS collects tracking data for over \numprint{40000} objects with a radius exceeding \numprint{10} cm and conducts an initial collision analysis \cite{SpaceForce}. Operators receive automatic collision warnings three times per day. However, as operators possess additional information about their satellites' orbit and planned maneuvers, they reperform the analysis with more precise data for their satellites. Note that in practice, the approach outlined in Section \ref{BackgroundCollisionRisk} is not employed in isolation; rather, multiple methodologies are applied concurrently to provide a more comprehensive understanding of the data (compare \cite{AidaKirschner}).\\[-.6em]
	
	Calculating the trajectory of the second object depends on observational data, which is an ongoing challenge. Various approaches have been developed based on different underlying assumptions (e.g., \cite{TLE_Orbit} and \cite{TLE_Orbit_1}). Despite progress in this research area, prediction accuracy is compromised by approximation errors and, most significantly, the lack of maneuver knowledge. Access to operator-provided data for predicting satellite movement would eliminate this error source, enhancing the accuracy of collision risk analysis.

	\section{Background on Fully Homomorphic Encryption}\label{BackgroundFHE}
	
	The core contribution of this work is the introduction of a novel model for collision risk analysis, leveraging the principles of FHE. In the classical paradigm, FHE is a single-user model that enables computations to be performed directly on encrypted data without the need for decryption, thereby preserving the confidentiality of the underlying information. The encryption process, denoted as $Enc$, is performed in a structure-preserving manner, such that for any function $f$ applied to the encrypted data, the result is equal to the encryption of the evaluation of $f$ on the plaintext data. As a prominent example, note that if $f$ is the addition or multiplication of two arguments, an FHE scheme fulfills
	\begin{align*}
		Enc(x) + Enc(y) &= Enc(x+y) \\
		Enc(x) \cdot Enc(y) &= Enc(x\cdot y).
	\end{align*}
	Note that it is sufficient to ask for the ability to homomorphically perform an arbitrary number of additions and multiplications, since more complex functions can be expressed through these two basic operations.\\[-.6em]
	
	The possibility to calculate on encrypted data enables the transfer of computationally intensive calculations to a non-trusted server without compromising the confidentiality of the data or the result of the calculation: The user hands over the encrypted data and the function to be performed on the data. The server carries out the computation on the encrypted data and returns the result to the user in an encrypted form. Holding the secret key, the user is the only one able to access the results of the computation through decryption. \\[-.6em]
	
\begin{figure}[ht]
		\centering
		\resizebox{\linewidth}{!}{%
	        \tikzset{
	messagebox1/.style={draw=red, minimum height=.5cm, anchor=west},
	errorbox1/.style={fill=red, draw=red, minimum height=.5cm, anchor=west, inner sep=0cm},
	messagebox2/.style={draw=blue, minimum height=.5cm, anchor=west},
	errorbox2/.style={fill=blue, draw=blue, minimum height=.5cm, anchor=west, inner sep=0cm},
	textsymbol/.style={minimum height=.5cm},
	brace/.style={decorate, decoration={brace, amplitude=7pt}},
	mirroredbrace/.style={decorate, decoration={brace, mirror, amplitude=7pt}},
	separator/.style={dashed},
}

\begin{tikzpicture}
	\draw[separator] (-0.35, 1.3) -- (-0.35, -5.3);
	\draw[separator] (-2.5, 1.3) -- (10.6, 1.3);
	
	
	\node[messagebox1, minimum width=1cm] (m1) at (0, 0) {$m_1$};
	\node[errorbox1, minimum width=0.4cm] (m1error) at (1, 0) {};
	\node[messagebox1, minimum width=1.8cm] (m1end) at (1.4, 0) {};
	
	\node[messagebox1, minimum width=1.1cm] (m2) at (3.7, 0) {$m_2$};
	\node[errorbox1, minimum width=0.6cm] (m2error) at (4.8, 0) {};
	\node[messagebox1, minimum width=1.5cm] (m2end) at (5.4, 0) {};
	
	\node[messagebox1, minimum width=1.5cm] (m1m2) at (7.4, 0) {$m_1 m_2$};
	\node[errorbox1, minimum width=1.7cm] (m1m2error) at (8.9, 0) {};
	
	\draw[brace] (0, 0.3) -- (1.4, 0.3);
	\node[textsymbol] (c1) at (0.7, 0.8) {$c_1$};
	\draw[brace] (3.7, 0.3) -- (5.4, 0.3);
	\node[textsymbol] (c2) at (4.55, 0.8) {$c_2$};
	\draw[brace] (7.4, 0.3) -- (10.6, 0.3);
	\node[textsymbol] (c1c2) at (9, 0.8) {$c_1 c_2$};
	
	\node[textsymbol] (cdot1) at (3.45, 0.8) {$\cdot$};
	\node[textsymbol] (equals1) at (7.15, 0.8) {$=$};
	\node[textsymbol] (cdot2) at (3.45, 0) {$\cdot$};
	\node[textsymbol] (equals2) at (7.15, 0) {$=$};
	
	\node[textsymbol, anchor=west, inner sep=0cm] (operation1) at (-2.5, 0.15) {Operation 1};
	\draw[separator] (-2.5, -1) -- (10.6, -1);
	
	
	\node[messagebox1, minimum width=1.45cm, minimum height=0.4cm, inner sep=0cm] (m1m2) at (0.05, -2) {$m_1 m_2$};
	\node[errorbox1, minimum width=1.65cm, minimum height=0.4cm] (m1m2error) at (1.5, -2) {};
	\node[messagebox2, minimum width=3.2cm] (bs1) at (0, -2) {};
	\node[errorbox2, minimum width=0.15cm] (bs1error) at (3.2, -2) {};
	\node[messagebox2, minimum width=1.55cm] (bs1end) at (3.35, -2) {};
	
	\node[messagebox2, minimum width=1.5cm] (bs2) at (7.4, -2) {$m_1 m_2$};
	\node[errorbox2, minimum width=0.15cm] (bs2error) at (8.9, -2) {};
	\node[messagebox2, minimum width=1.55cm] (bs2end) at (9.05, -2) {};
	
	\node[textsymbol, anchor=west, inner sep=0cm] (bootstrapping) at (-2.5, -2) {Bootstrapping};
	\draw[->] (5.2, -2) -- (7.1, -2);
	\draw[separator] (-2.5, -3) -- (10.6, -3);
	
	
	\node[messagebox2, minimum width=1.5cm] (m1m2fresh) at (0, -4) {$m_1 m_2$};
	\node[errorbox2, minimum width=0.15cm] (m1m2fresherror) at (1.5, -4) {};
	\node[messagebox2, minimum width=1.55cm] (m1m2freshend) at (1.65, -4) {};
	
	\node[messagebox2, minimum width=1.1cm] (m3) at (3.7, -4) {$m_3$};
	\node[errorbox2, minimum width=0.3cm] (m3error) at (4.8, -4) {};
	\node[messagebox2, minimum width=1.8cm] (m3end) at (5.1, -4) {};
	
	\node[messagebox2, minimum width=1.7cm, inner sep=0cm] (m1m2m3) at (7.4, -4) {\small $m_1 m_2\hspace{-.07cm}+\hspace{-.06cm}m_3$};
	\node[errorbox2, minimum width=0.35cm] (m1m2m3error) at (9.1, -4) {};
	\node[messagebox2, minimum width=1.15cm] (m1m2m3end) at (9.45, -4) {};
	
	\draw[mirroredbrace] (0, -4.3) -- (1.65, -4.3);
	\node[textsymbol] (c1) at (0.825, -4.8) {$c_1 c_2$};
	\draw[mirroredbrace] (3.7, -4.3) -- (5.1, -4.3);
	\node[textsymbol] (c2) at (4.4, -4.8) {$c_3$};
	\draw[mirroredbrace] (7.4, -4.3) -- (9.45, -4.3);
	\node[textsymbol] (c1c2) at (8.425, -4.8) {$c_1 c_2 + c_3$};
	
	\node[textsymbol] (plus1) at (3.45, -4) {$+$};
	\node[textsymbol] (equals3) at (7.15, -4) {$=$};
	\node[textsymbol] (plus2) at (3.45, -4.8) {$+$};
	\node[textsymbol] (equals4) at (7.15, -4.8) {$=$};
	
	\node[textsymbol, anchor=west, inner sep=0cm] (operation2) at (-2.5, -4.15) {Operation 2};
	\draw[separator] (-2.5, -5.3) -- (10.6, -5.3);
	

	\draw[->, rounded corners] (9, -0.35) -- (9, -0.6) -- (2.45,-1.3) -- (2.45, -1.6);
	\draw[->, rounded corners] (9, -2.35) -- (9, -2.6) -- (1.6,-3.3) -- (1.6, -3.6);
\end{tikzpicture}
		}
		\caption{Visualization of addition and mulitplication in somewhat homomophic schemes as well as the bootstrapping technique for noise reduction.}
		\label{Bootstrapping}
	\end{figure}
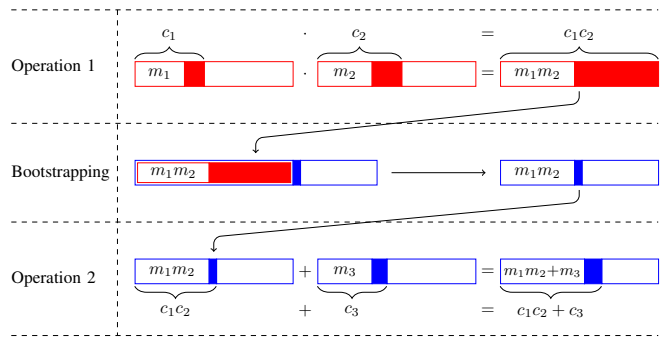

	The concept of FHE was first proposed by Rivest, Adleman, and Dertouzos in 1978 \cite{Rivest_Adleman_Dertouzos}, based on their observation that for the RSA cryptosystem \cite{RSA}, it holds that
	\begin{align*}
		Enc(x)\cdot Enc(y) = Enc(x\cdot y).
	\end{align*}
	While often not explicitly designed with this capability, several encryption schemes naturally support either homomorphic multiplication or addition, such as the ElGamal \cite{ElGamal} or the Paillier scheme \cite{Paillier}. However, no encryption system supporting an unlimited number of both operations was found until the groundbreaking work of Gentry in 2009 \cite{Gentry}. Gentry started from a so-called somewhat homomorphic scheme that supports a limited number of additions and multiplications. Such schemes typically encrypt a message by encoding it in an appropriate manner, such as representing it as a point on a lattice, and then adding a noise or error term. Consequently, the addition of ciphertexts involves summing the noise terms, while the multiplication of ciphertexts results in an even more exacerbated noise. Repeating this process multiple times yields a scenario where the receiver is unable to decrypt the message due to the high noise, thus constraining the number of permissible additions and multiplications. It is important to note that while symmetric variants of these schemes do exist, the majority of the current somewhat homomorphic schemes are public-key based. This means that the user generates a key pair comprising a public key $pk$ and a secret key $sk$. While anyone can employ the public key to encrypt a message, only the user, with the secret knowledge of $sk$, can decrypt messages. On that basis, Gentry introduced a novel technique, known as bootstrapping, to mitigate the issue of growing noise in somewhat homomorphic schemes. Upon detecting that a predefined noise threshold has been reached, a bootstrapping step is executed, which starts with the generation of a new key pair $(pk', sk')$. The server utilizes the newly generated public key $pk'$ to encrypt the current ciphertext, thereby resulting in the ciphertext being doubly encrypted. Subsequently, the user provides the encryption of the original secret key $sk$ under $pk'$. This action allows the server to remove the underlying encryption layer, leaving only the encryption with the new public key $pk'$ intact. This method ensures the preservation of the privacy of both the secret key and the data, while effectively reducing the noise associated with the system. A visual description of the process is depicted in Figure \ref{Bootstrapping}. \\[-.6em]
	
	That way, a somewhat homomorphic scheme can be extended to a fully homomorphic scheme. However, a crucial prerequisite is that the underlying somewhat homomorphic scheme supports more additions and multiplications than are required by the decryption operations. Otherwise, the bootstrapping step becomes unattainable.\\[-.6em]
	
	Every bootstrapping step is time-consuming, particularly affecting the performance of early schemes and resulting in prohibitively large computation times. However, advancements within the field of somewhat homomorphic schemes have led to the development of more efficient schemes that support a larger number of additions and multiplications, thereby reducing the frequency of bootstrapping steps. Additionally, the emergence of more effective bootstrapping methods has further improved the process. Although the overhead remains significant, the combined advancement of somewhat homomorphic schemes and more effective bootstrapping methods brings real-world applications within reach.

	\section{Multi-party Fully Homomorphic Encryption}
 
	In our specific application, a traditional FHE scheme is not applicable. Instead, we are facing a scenario where multiple operators wish to perform calculations on their joined data sets on a shared server without disclosing their information to one another or the server. To address this challenge, mainly two distinct multi-party FHE frameworks exist. \\[-.6em]
	
	In threshold FHE (compare, e.g. \cite{ThresholdFHE}, \cite{ThresholdFHE1}) the parties collectively generate a shared public key and a secret key, which is distributed among all participants. Each party can encrypt their respective data using the public key. Consequently, the server processes the calculations on data encrypted under the same key. Since each party holds a share of the secret key, decryption requires the participation of a certain number of parties, known as the threshold. In our scenario, to maintain privacy in all cases, the full-threshold scenario is employed, where all parties must participate in the decryption process. It is important to note that adding a new party to a threshold FHE scheme is resource-intensive, as it necessitates a new key setup for all parties. However, due to the shared key, the computational cost on the server side is comparable to a single-party scenario, making this framework particularly appealing for complex calculations or scenarios with many parties.  \\[-.6em]
 
	Another approach to achieve multi-party FHE is the utilization of multi-key FHE (see, e.g. \cite{On-the-flyMultiparty}, \cite{Lopez-AltTromerEtAl2017MultikeyFullyHomomorphic}). In this framework,  each operator independently generates an encryption key and encrypts the own data accordingly. In this setup, the server must be capable of handling ciphertexts encrypted with different keys and still accurately compute the collision risk. Notably, decryption must be performed collectively by all parties involved. One advantage of the multi-key framework is that the theoretical number of participants is unlimited, allowing the inclusion of all satellite operators within a common framework where computations are performed pairwise. Integration of new parties is also straightforward. However, while the multi-key scenario is manageable from the client-side, the computational burden for the server increases polynomially with the number of parties involved \cite{ChenDaiKimSong}. A high-level summary of the differences between both types of multi-party FHE schemes can be found in Table \ref{Multiparty}.\\[-.6em]
 
	\begin{table}
		\caption{High-level comparison of multi-key FHE and threshold FHE.}
		\centering
		\begin{tabular}{@{}lll@{}}
			\toprule
			& Multi-key FHE           & Threshold FHE      \\
			\midrule
			\midrule
			Encryption type      & symmetric or asymmetric & asymmetric         \\ \midrule
			Key generation       & independent             & collaborative      \\ \midrule
			Encryption           & under different keys    & under the same key \\ \midrule
			Decryption           & collaborative           & collaborative      \\ \midrule
			Integration of       & easy                    & costly             \\
			new parties          &                         &                    \\ \midrule
			Server-sided         & growing with            & constant           \\
			computational effort & number of parties       &                    \\
			\bottomrule
		\end{tabular}
		\vspace{0.1cm}
		\label{Multiparty}
	\end{table}

	\section{Fully Homomorphic Encryption for Satellite Collision}\label{Fully_Homomorphic_Encryption_for_Satellites}
	
	One of the most significant hurdles impeding the widespread usage of FHE schemes is the high computational complexity, which results in substantial computation times. In this regard, threshold FHE offers a more attractive alternative for most scenarios, particularly as the number of participating parties increases. As a consequence, threshold FHE schemes have emerged as the preferred variant for multi-party FHE applications, and as such, will be the primary focus of the following discussion. However, it is worth noting that multi-key FHE schemes remain an active area of research, and given that our application only requires computations to be performed between two parties, it may serve as a use case for multi-key schemes as well.  \\[-.6em]
 
	One of the inherent features of full-threshold FHE schemes is the collaborative decryption. However, this necessity poses a significant threat to the overall system reliability, particularly in scenarios where one or more parties become unreachable. To mitigate this risk, we employ a pairwise approach, leveraging the inherently pairwise nature of satellite collision risk analysis. This includes generating a full-threshold FHE scheme between each pair of operators, thereby resulting in a more resilient system that is better equipped to handle the failure of individual parties. Furthermore, this approach simplifies the integration of new parties, as the existing pairwise relationships remain unaffected.\\[-.6em]
	
	\begin{Remark}
		In contrast to the threshold FHE setting, the multi-key scenario inherently lends itself to a pairwise approach. Typically only the parties that actively participated in the computation need to be involved in the decryption process. Hence, multi-key FHE schemes exhibit inherent resilience against failures of single parties. \\[-.6em]
	\end{Remark}
 
	Most threshold FHE schemes are restricted to processing bit or integer inputs and are unable to handle the real values required for calculating the collision probability integral. We can address this limitation by scaling all real numbers by a suitable factor, performing the calculations on the resulting integers, and then rescaling the results. While this approach ensures that the calculations can be executed homomorphically, it simultaneously restricts the precision that can be achieved.\\[-.6em]
	
	To avoid manual scaling, we consider the CKKS (Cheon--Kim--Kim--Song) scheme \cite{CKKS} in a threshold scenario. The CKKS scheme operates on complex and, consequently, on real numbers, making it suitable for our application without further manual modifications. To handle complex and real numbers, CKKS employs approximate arithmetic, which involves performing operations on the real numbers with a small, controlled approximation error. By facilitating bootstrapping, CKKS enables an unlimited number of operations, as outlined in Section \ref{BackgroundFHE}. \\[-.6em]
	
	Using the implementation of threshold CKKS in OpenFHE \cite{OpenFHE}, we construct a threshold setting, in which we aim to calculate the collision risk. As input values, we obtain the radii as well as the covariance matrices from two operators as described in Section \ref{BackgroundCollisionRisk}. For simplification, we assume that the data has been previously transformed into the encounter-plane coordinate system, yielding values for $r$, $\sigma_{x'}$, and $\sigma_{z'}$. Each operator is presumed to possess an additive share of these values. We now approximate the collision probability integral in (\ref{Approx}) using the different integration rules presented above. \\[-.6em]
	
	Given that homomorphic schemes, by default, support only additions and multiplications, the evaluation of the function $p(y,\phi)$ in (\ref{Approx}), including the exponential function, inverses, and trigonometric functions, presents a significant challenge. Hence, as an initial step, we employ a precomputation strategy utilizing an encrypted lookup table. Specifically, during a preliminary calculation phase, we generate tables containing encrypted function evaluations $p(y,\phi)$ for various values of $\sigma_{x'}$ and $\sigma_{z'}$. These tables are then stored in conjunction with the encrypted values of $\sigma_{x'}$ and $\sigma_{z'}$. \\[-.6em]
 
	Upon initiation of a collision probability analysis, the encrypted parameters for the true covariances need to be compared against the available table parameters. However, direct comparison within homomorphic schemes is not possible. Several approaches exist to enable secure comparison in homomorphic settings (see, e.g., \cite{ComparisonFH}, \cite{ComparisonFH1}). However, a comprehensive examination of these approaches is beyond the scope of the present paper. Hence, for the purposes of this discussion, we presuppose the ability to identify the correct table.
	When utilizing a precomputed table to evaluate the integral, the requisite operation is merely the addition of encrypted values. Notably, this process is significantly more efficient than multiplication, owing to the reduced rate of noise growth. If we isolate this step and consider it separately, it becomes apparent that even a partially homomorphic scheme would suffice for the calculation. In fact, alternative approaches may be more optimal for this specific step. However, as our ultimate objective is to explore the feasibility of fully homomorphic computation of the integral, it is necessary to employ a fully homomorphic scheme like CKKS at this juncture.  \\[-.6em]
	
	The implemented CKKS scheme was initialized with a security level of \numprint{128} bits. A notable feature of the CKKS scheme is its flexibility in allowing the user to specify the multiplicative depth, which represents the maximum number of consecutive multiplications that can be performed before the accumulation of noise renders decryption infeasible. From a mathematical perspective, as the multiplicative depth increases, the dimension of the underlying ring increases simultaneously making every operation increasingly costly. In the context of utilizing precomputed tables, no multiplicative depth is required and the scheme is computationally efficient, as illustrated in Table \ref{Numerical_Approximation} for the exemplary values $r=5$, $\sigma_{x'}=50$ and $\sigma_{z'}=25$ (in meters). The calculations leverage the inherent feature of CKKS to perform computations within multiple slots simultaneously, thereby minimizing the number of required additions. Notably, for all parameter choices, the costly bootstrapping step is not used, which further contributes to the scheme's efficiency. \\[-.6em]
	
	\begin{table*}
		\centering
		\caption{Approximation of the collision probability $P_{col}$ with the Threshold CKKS scheme with different numerical integration rules and step sizes for $r=5$, $\sigma_{x'}=50$, and $\sigma_{z'}=25$ (in meters). The exact function evaluations are precomputed and stored in lookup tables. The number of additions ($+$) is recorded for a non-parallel computation for comparisons with other schemes.}
		\begin{tabular}{@{}lllrrrrr@{}}%
			\toprule
			\multirow{2}{*}{Integration rule} & \multirow{2}{*}{$h_r$} & \multirow{2}{*}{$h_{\phi}$} & \multirow{2}{*}{Abs. error} & Rel. error  & Time  & \# of function & \multirow{2}{*}{\# of  $+$ } \\
			&                        &                             &                             & (in \%)              & (in s) & evaluations    &                            \\
			\midrule
			\midrule
			Trapezoidal                       & 0.5                    & 0.5                         & $4.69\cdot 10^{-7}$         & $4.94\cdot 10^{-3}$  & 0.16   & 143            & 480                        \\
			& 0.1                    & 0.1                         & $2.29\cdot 10^{-8}$         & $2.33\cdot 10^{-4}$  & 0.16   & 3,213          & 12,400                     \\
			& 0.05                   & 0.05                        & $6.00\cdot 10^{-9}$         & $6.06\cdot 10^{-5}$  & 0.21   & 12,726         & 50,000                     \\
			\midrule
			Trapezoidal + Simpson & 0.5 & 0.5
			& $6.06\cdot 10^{-7}$ & $6.39\cdot 10^{-3}$ & 0.16 & 286 & 1,200 \\
			& 0.1                    & 0.1                         & $2.46\cdot 10^{-8}$         & $2.50\cdot 10^{-4} $ & 0.17   & 6,426          & 31,000                     \\
			& 0.05                   & 0.05                        & $ 6.19\cdot  10^{-9}$         & $6.23\cdot 10^{-5}$  & 0.17   & 25,452         & 125,000                    \\
			\midrule
			Gaussian quadrature (order $2$)   & 0.5                    & 0.5                         & $7.18\cdot 10^{-11}$        & $7.56\cdot 10^{-7}$ & 0.16 & 480 & 480 \\
			& 0.1                    & 0.1                         & $1.51\cdot 10^{-11}$        & $1.54\cdot 10^{-7}$  & 0.16   & 12,400         & 12,400                     \\
			& 0.05                   & 0.05                        & $4.01\cdot 10^{-12}$        & $4.05\cdot 10^{-8}$  & 0.17   & 50,000         & 50,000                     \\
			\midrule
			Gaussian quadrature (order $3$)   & 0.5                    & 0.5                         & $9.20\cdot 10^{-13}$        & $9.69\cdot 10^{-9}$ & 0.16 & 1,080 & 1,080  \\
			& 0.1                    & 0.1                         & $1.89\cdot 10^{-14}$        & $1.93\cdot 10^{-10}$ & 0.16   & 27,900         & 27,900                     \\
			& 0.05                   & 0.05                        & $1.39\cdot 10^{-14}$        & $1.41\cdot 10^{-10}$ & 0.19   & 112,500        & 112,500                    \\
			\midrule
			Gaussian quadrature (order $4$)   & 0.5                    & 0.5                         & $4.21\cdot 10^{-14}$        & $4.44\cdot 10^{-10}$  & 0.16 & 1,920 & 1,920 \\
			& 0.1                    & 0.1                         & $1.99\cdot 10^{-14}$        & $2.03\cdot 10^{-10}$ & 0.22   & 49,600         & 49,600                     \\
			& 0.05                   & 0.05                        & $1.49\cdot 10^{-14}$        & $1.51\cdot 10^{-10}$ & 0.24   & 200,000        & 200,000                    \\
			\bottomrule
		\end{tabular}
		\label{Numerical_Approximation}
	\end{table*}
 
	As illustrated in the table, for the selected input values and parameters, the trapezoidal rule requires the fewest function evaluations, yet its convergence rate is slower compared to the Gaussian quadrature. Notably, the implementation of Simpson's rule does not yield significant improvements, and considering the associated computational cost, it is not a viable option. Depending on the desired level of precision, we recommend employing a Gaussian quadrature of order 2 or 3 with a larger step size as a compromise between accuracy and the number of function evaluations. However, due to the parallel computation, all numerical integration rules can be executed within a reasonable time frame. It is essential to acknowledge that the CKKS scheme is an approximate method, which inherently limits the achievable precision of the numerical integral rules. Furthermore, the parallel calculation may introduce scaling errors, as all slots are scaled uniformly. These limitations should be taken into account when selecting and implementing a CKKS scheme. \\[-.6em]
 
	The approach of utilizing precalculated tables is beset by several drawbacks. Firstly, the requirement to securely compare table parameters poses a significant challenge within the homomorphic setting. Additionally, reliance on pre-existing tables can introduce considerable errors if the actual values of $\sigma_{x'}$ and $\sigma_{z'}$ do not precisely align with the table's parameters, necessitating the selection of the best-fitting table instead. Moreover, the storage of these precalculated tables demands substantial memory resources.\\[-.6em]
 
	Therefore, in a second step, we evaluate the possibility of calculating the function evaluation homomorphically in real time using the encrypted values of $r$, $\sigma_{x'}$, and $\sigma_{z'}$ instead of precomputed tables. Provided that the computation can be performed within a reasonable time frame, this approach is expected to provide a more feasible solution for practical applications. \\[-.6em]
	
	When the overall number of multiplications required for a specific computation exceeds the multiplicative depth of the scheme, bootstrapping has to be employed at least once during the computation. Note that each bootstrapping step itself contains multiplications and thus needs to be triggered in time such that its application does not exceed the multiplicative depth of the scheme. For example, the bootstrapping step in the CKKS scheme requires 16 multiplications and a multiplicative depth of 26 thus effectively allows for 10 multiplications on the encrypted data at a time. Choosing the multiplicative depth is generally a trade-off between overall speed of the scheme and the number of bootstrapping steps required. Without claiming to be optimal, we have chosen a depth of 26 for our implementation.
	Notably, this parameter adjustment has a profound impact on the scheme's performance, as evident in Table \ref{Comp_Run_Time}, which compares the run times of the CKKS scheme using the Gaussian quadrature of order 2 with precomputed function evaluations for both depths. The data reveals a substantial slowdown in the scheme's execution time, underscoring the significant computational overhead associated with increased multiplicative depth, or equivalently, increased dimension of the underlying ring. \\[-.6em]
	
	\begin{table}
		\centering
		\caption{CKKS run time for the Gaussian quadrature rule of order 2 with different multiplicative depths.}
		\begin{tabular}{@{}llrr@{}}%
			\toprule
			\multirow{3}{*}{$h_r$} & \multirow{3}{*}{$h_{\phi}$} & Time (in s)       & Time (in s)        \\
			&                             & (mult. depth = 0) & (mult. depth = 26) \\
			\midrule
			\midrule
			0.5                    & 0.5                         & 0.16              & 213.16             \\
			0.1                    & 0.1                         & 0.16              & 215.62             \\
			0.05                   & 0.05                        & 0.17              & 227.12             \\
			\bottomrule
		\end{tabular}
		\label{Comp_Run_Time}
	\end{table}
	
	On top of that, the function evaluations contribute massively to the computation time. As indicated in equation (\ref{Approx}), the function $p$ to be evaluated entails the calculation of inverses for $\sigma_{x'}$ and $\sigma_{z'}$, as well as their squared values. Moreover, it necessitates the evaluation of cosine, sine, and exponential functions. Given that within fully homomorphic encryption schemes, every function needs to be expressed in terms of additions and multiplications, computing these components poses a significant challenge. \\[-.6em]
 
	To illustrate the computational complexities and time requirements associated with such calculations, we perform point evaluations of $p$ utilizing the CKKS scheme with depth $26$. For the sake of simplicity, we assume that the inverses of $\sigma_{x'}$, $\sigma_{z'}$, and their squared values are known precisely without further computation, which could, for example, be realized by lookup tables. We subsequently approximate the exponential function and the cosine function via their respective Taylor series expansions, given by
	\begin{align*}
		\exp(x)=\sum_{n=0}^{N_1} \frac{x^n}{n!} \text{ and } \cos(x)=\sum_{n=0}^{N_2} (-1)^n \frac{x^{2n}}{(2n+1)!}
	\end{align*}
	for approximation orders $N_1, N_2\in\mathbb{N}$. Table \ref{Function_Eval} presents the calculated error, computational time, and number of bootstrapping steps for the single point evaluation $p(y,\phi)$, where $y=5$, $\phi=2\pi$, $\sigma_{x'}=50$, and $\sigma_{z'}=25$ (in meters). The results indicate that the precision of the calculation is inherently limited by the parameters of the CKKS homomorphic encryption scheme. Specifically, augmenting the multiplicative depth to achieve higher precision comes at the cost of increased computational time. It is worth noting that the results presented in the table pertain to a single point evaluation. However, the CKKS scheme does offer the possibility of parallelizing computations across multiple slots, which could potentially accelerate the process. Nevertheless, this parallelization also introduces an additional, non-negligible error due to the scaling operation being performed across all slots simultaneously. \\[-.6em]
	
	\begin{table}
		\centering
		\caption{Homomorphic approximation of a single function evaluation $p(y,\phi)$ using CKKS for approximation orders $N_1$ and $N_2$ with parameters $y=5$, $\sigma_{x'}=50$, $\sigma_{z'}=25$ (in meters), and $\phi=2\pi$. We list run time, number of bootstrapping steps (BS), absolute and relative error, and number of additions ($+$) and multiplications ($\cdot$).}
		\begin{tabular}{@{}rrrrlrrr@{}}%
			\toprule
			\multirow{2}{*}{$N_1$} & \multirow{2}{*}{$N_2$} & Time     & \# of & \multirow{2}{*}{Abs. error} & Rel. error & \# of \\
			&                        & (in s)   & BS    &                             & (in \%)             & $+$ / $\cdot$ \\
			\midrule
			\midrule
			5                      & 5                      & 363.79   & 4     & $3.39\cdot 10^{-4}$         & 53.51               & 12 / 30       \\
			5                      & 10                     & 449.02   & 5     & $4.80\cdot 10^{-9} $        & $7.58\cdot 10^{-4}$ & 17 / 40       \\
			10                     & 10                     & 650.64   & 7     & $5.48\cdot 10^{-10}$        & $8.65\cdot 10^{-5}$ & 22 / 50       \\
			10                     & 15                     & 2,439.25 & 25    & $5.58\cdot 10^{-8}$         & $8.81\cdot 10^{-3}$ & 27 / 60       \\
			\bottomrule
		\end{tabular}
		\label{Function_Eval}
	\end{table}
 
	As indicated in \cite{Chan}, the relative error of the approximation of the collision probability by to the method of Foster and Estes in use is in the range of $10^{-6}$ percent. If we aim to reach a similar level with homomorphically encrypted calculations, precise calculation of the function evaluations is necessary. But as shown above, this requires a large multiplicative depth, which in turn results in prohibitively large computation times. Furthermore, our model still incorporates certain simplifications that would need to be addressed in real-world applications. In an operational context, frequent updates of the collision probability are necessary, underscoring the need for an efficient and accurate computation method. Despite its efficiency, the implemented scheme is not suitable to fulfill all requirements simultaneously.\\[-.6em]
	
	As mentioned before, the collision risk approximation by Foster and Ester that we are basing our work on is just one possible solution -- other methods or representations might be more suitable for homomorphic calculations. Depending on the requirements for precision and run time, other schemes with their own strengths and weaknesses may also be considered. With the rapid development of more efficient schemes and bootstrapping procedures, we see great potential for future use of FHE in satellite collision risk analysis. As a result, the increasing risk of satellite collisions, associated with a potential satellite loss and an increasing amount of space debris, can be minimized in a privacy-preserving manner.
 
	\bibliographystyle{IEEEtran}
	\bibliography{references}

\end{document}